\documentclass[aps,prl,twocolumn,longbibliography]{revtex4-1}
\usepackage{natbib,graphicx,dcolumn,color,lipsum,hyperref}
\usepackage{amsmath,amssymb,verbatim,moreverb,bm,mdframed}
\usepackage{siunitx}

\def\bef{\begin{mdframed}}
\def\eef{\end{mdframed}}
\def\be{\begin{equation}}
\def\ee{\end{equation}}
\def\ber{\begin{eqnarray}}
\def\eer{\end{eqnarray}}

\def\bsigma{\mbox{\boldmath $\sigma$}}

\def\br{{\bf r}}

\def\bhx{{\bf \hat x}}

\def\bp{{\bf p}}
\def\bV{{\bf V}}

\def\bb{{\bf b}}
\def\bB{{\bf B}}
\def\bM{{\bf M}}
\def\bH{{\bf H}}

\def\bS{{\bf S}}
\def\nn{\nonumber}

\def\Hcal{{\mathcal H}}

\newcommand\commentout[1]{}

\def\barS{{\bar{S}}}
\def\barbS{{\bar{\bf S}}}

\newcommand{\ie}{{\it i.e.~}} 	
\newcommand{\eg}{{\it e.g.~}} 	

\def\bDelta{\mbox{\boldmath $\Delta$}}
\def\BB{{\mathbb B}}
\def\cR{{\mathcal R}}

\def\Mv{{\bf M}}

\def\wv{{\bf w}}
\def\mv{{\bf m}}
\def\xv{{\bf x}}

\def\Rv{{\bf R}}

\def\Hv{{\bf H}}
\def\Iv{{\bf I}}
\def\Mv{{\bf M}}

\def\nn{\nonumber}

\def\chiv{\bm{\chi}}

\def\psiv{\bm{\psi}}
\def\sigmav{\bm{\sigma}}

\def\Hparallel{{H_\parallel}}
\def\BB{{\mathbb B}}

\definecolor{greenS}{rgb}{0.00, 0.6, 0.00}
\definecolor{orangeS}{rgb}{0.6, 0.1, 0.1}

\begin{document}
 \title{Topological fragility and bilinear magnetoelectric resistance in gapless edge states}
 \author{Cosimo Gorini$^1$, Matthieu Bard$^2$, Sophie Gueron$^2$, Hélène Bouchiat$^2$ Giovanni Vignale$^3$}
 
 \affiliation{$^1$Université Paris-Saclay, CEA, CNRS, SPEC, 91191 Gif-sur-Yvette, France \\
 $^2$Université Paris-Saclay, CNRS, Laboratoire de Physique des Solides, 91405 Orsay, France \\
 $^3$The Institute for Functional Intelligent Materials (I-FIM), National University of Singapore, 4 Science Drive 2, Singapore 117544
 }

\date{\today}

\begin{abstract}   
In time-reversal symmetric systems such as topological and higher-order topological insulators, 1D spin-momentum locked edge and hinge states are theoretically ``perfectly conducting'', being immune to backscattering by non-magnetic disorder. Here, we reveal a fundamental ``topological fragility'': these states exhibit a bilinear magnetoelectric resistance significantly larger than in 2D systems. This effect requires two ingredients: (i) spin-momentum locking, which maximizes time-reversal symmetry breaking in the non-linear regime, and (ii) random spin-orbit interaction -- the same mechanism behind Elliott - Yafet spin relaxation in heavy elements. Together, these generate a robust backscattering channel when a modest external magnetic field is applied. Our theory requires no gap opening or complex many-body effects, offering a simple and general mechanism that quantitatively explains recent observations in Bismuth hinge states.
\end{abstract}
\maketitle


Topologically insulating phases host conducting boundary states where electronic spin and momentum are strictly locked, forming Dirac-type dispersions at low energy \cite{hasan2010,schindler2018a,schindler2018b,geier2018}. Frontier states exhibit unusual nonlinear transport phenomena, where resistance depends on both the magnitude and direction of current and magnetic field \cite{he2018,he2019,dyrdal2020,zhao2020,zhao2021}. A prime example is bilinear magnetoelectric resistance (BMER) on the surface of three-dimensional (3D) topological insulators, \eg ${\rm Bi}_2{\rm Se}_3$ \cite{he2018} 
\be 
\label{eq_BMER_0} 
R=R^{(0)}+A I H 
\ee 
where $R^{(0)}$ is the linear electric resistance, $I$ is the current, $H$ the magnetizing field, and $A$ a constant dependent on the orientation of $I$ and $H$ with respect to the crystallographic axes.

Two mechanisms for BMER have been identified. The first \cite{he2018,he2019} arises from the second-order response of a warped Fermi surface to an electric field; it vanishes if the surface is perfectly circular. The second \cite{dyrdal2020} relies on the scattering rate dependence on current-induced spin polarizations. Both require at least two dimensions: the first due to transverse warping, the second due to scattering events driven by Dirac cone fluctuations transverse to transport -- a degree of freedom absent in one dimension.

\begin{figure}
\includegraphics[width=0.95\columnwidth]{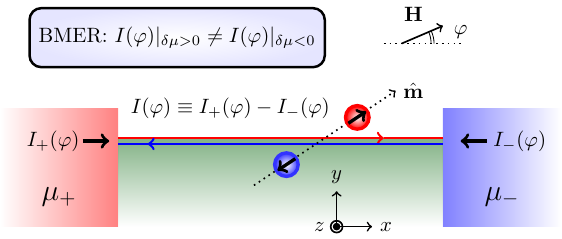}
\caption{Schematics of a spin-momentum locked 1D  wire connecting reservoirs at  chemical potentials $\mu_+$,$\mu_-$. The magnetizing field $\Hv$ forms an angle $\varphi$ with the channel, while the electronic spins are parallel to the spin-momentum locking direction $\hat{\bf m}$. The net current $I(\varphi)$ is different under positive ($\delta\mu>0$) and negative  ($\delta\mu<0$) bias, with $\delta\mu\equiv\mu_+-\mu-$.  The asymmetry is quantified by the resistance $R_{BMER}$, Eq.~\eqref{eq_BMER_only1}.}
\label{fig_intro}
\end{figure}

Neither option is available for one-dimensional (1D) frontier states (edge or hinge states), where BMER manifests as non-reciprocal transport (see Fig.~\ref{fig_intro}). Yet such an effect was recently observed and used to demonstrate spin-momentum locking \cite{zhao2021}. This presents a paradox, as 1D spin-momentum locked states are theoretically robust against the backscattering required for BMER \cite{wu2006,hasan2010}. Previous proposals invoked complex extrinsic mechanisms: coupling to bulk midgap states with Hubbard interactions \cite{chen2024}, inelastic many-body scattering \cite{schmidt2012}, two-particle processes such as Umklapp \cite{wu2006}, or gap opening by $e$-$e$ interactions combined with non-linear spectral distortions \cite{balram2019}. The fundamental question remains: is there a simpler, ubiquitous mechanism enabling backscattering and thus causing BMER in topologically protected 1D Dirac states?

Here we demonstrate that the answer is ``yes''. The critical observation is that strong spin-orbit coupling is not only a typical prerequisite of topological protection \cite{hasan2010}, but also the solution to the backscattering problem when $H\neq0$. In high-$Z$ materials spin-orbit coupling may lead to band-inversion and spin-momentum locking \cite{hasan2010}, while its fluctuations due to ambient disorder ensure momentum and spin relaxation rates are of the same order of magnitude \cite{meservey1978,komnik2005}. This suffices to generate a non-reciprocal, second-order transport correction ($A I H$) to the quantized linear resistance. Unlike previous proposals, our mechanism requires no inelastic processes, gap opening, strong many-body effects, or bulk coupling.

The key point in our analysis is that spin-momentum locking, while guaranteeing perfect linear-response transmission in the absence of a magnetic field, maximizes backscattering by breaking time-reversal symmetry in the {\it non-linear} regime. The current-carrying state generates a ``perfect'' spin polarization, where each charge carrier contributes to the spin along the locking axis. There is no partial cancellation from other bands (as in Rashba 2D systems) nor losses from diffusive dynamics possible in 2D. This yields an effective magnetic moment that maximally breaks time-reversal symmetry without opening a gap.

There are, in principle, two time-reversal symmetry-breaking mechanisms: (i) the Fock exchange mean field from Coulomb interactions \cite{chen2024,balram2019}, and (ii) the field from magnetization via classical magnetostatics. We focus on (i), dominant by several orders of magnitude \cite{SuppMat}. Crucially, neither mechanism would produce BMER without a spin-flipping scattering channel. We show that random spin-orbit interaction provides such a channel via the Elliott-Yafet process. Using this mechanism, we obtain a theoretical estimate for 1D BMER that agrees excellently with experimental data for hinge states in Bismuth \cite{bardpreprint}. 

{\it BMER from linear magneto-resistance} -- Let us start with a general argument leading to BMER in 1D Dirac edges or hinges topologically protected against backscattering, see Fig.\ref{fig_intro}.  These are perfectly transmitted transport channels no matter the strength of non-magnetic or spin-orbit coupling disorder. The linear response electrical resistance of each state in such a two-terminal device is the contact resistance alone, $R^{(0)}=h/e^2$, with $e, h$ respectively the electronic charge and Planck constant.  What if a magnetic field $\Hv$ is switched on?  Since the field breaks time-reversal symmetry the resistance must increase, but according to Onsager’s reciprocity theorem it cannot change under reversal $\Hv\to-\Hv$. For sufficiently small magnetic fields one thus has
\be
\label{eq_BMER_1}
R^{(0)}(\Hv) = R^{(0)} + \frac{1}{2}\sum_{i,j=x,y,z}H_i R_{ij} H_j,
\ee
with $R_{ij}$ a matrix dependent on the microscopics of the problem -- Latin indices $i, j$ (later also $l$) denote here and throughout the 3D Cartesian coordinates $x, y, z$.  

We emphasize that Eq.~(\ref{eq_BMER_1}) is a {\it linear} resistance in the presence of a magnetic field: BMER is however a non-linear phenomenon.  Resistance contributions odd in $H$ and $I$ such as the one in Eq.~\eqref{eq_BMER_0} may in general appear beyond linear response \cite{sanchez2004,lofgren2004,spivak2004,angers2007}, and in particular if the effective magnetic field acting on the electrons depends on the current.  To this end, let us add to $\bH$ an internally generated magnetization field $\Mv$ -- whose origin, exchange or magnetostatics, is for the moment irrelevant.  Therefore $\bH \to \bH + \bM$, and one has
\be
\label{eq_BMER_2}
R(\Hv,\Mv)=R^{(0)}+\sum_{i,j} \left(H_i +M_i\right) R_{ij} \left(H_j +M_j\right).
\ee 
Under equilibrium conditions (excluding the case of spontaneous magnetization)  $\Mv \propto \Hv$ and the magneto-resistance is then quadratic in $H$.  The situation is different if the system is driven out of equilibrium by an electric field, for in this case the current $\Iv$ flowing in the non-equilibrium state creates a magnetization field  $M_i = \sum_l \chi_{il} I_l$ which must be added to $\Hv$ -- this is a form of current-induced spin polarization typical of spin-orbit coupled systems \cite{ivchenko1978,vorobev1979,aronov1989,edelstein1990,kato2004,gorini2017}. Expressing $\Mv$ in terms of $\Iv$ we  get
\be
\label{eq_BMER_derived}
R(\Hv,\Iv) = R^{(0)}+ \frac{1}{2}\sum_{i,j,l} \left(H_i + \chi_{il} I_l\right) R_{ij} \left(H_j + \chi_{jl} I_l\right).
\ee
The terms linear in $I$ and $H$ on the right hand side are responsible for BMER.  Comparing with Eq.~\eqref{eq_BMER_0} we identify 
\be
A= \sum_{i,j}\hat I_i [\chiv \cdot \Rv]_{ij}\hat H_j\,,
\ee
where $\chiv$ and $\Rv$ are matrices with components $\chi_{ij}$, $R_{ij}$ and $\hat \Iv$,$\hat \Hv$ are unit vectors with components $I_i/I$ and $H_i/H$ respectively.  We remark that other paths to BMER exist, \eg the quadratic current response in non-centrosymmetric systems \cite{he2019}, and we present a general Landauer-Büttiker formulation including both in the End Matter.  Eq.~\eqref{eq_BMER_derived} represents however the most direct and intuitive mechanism.  It is also the unique pathway available at low energies where deviations from a linear dispersion are negligible, providing a natural explanation for BMER in 1D topological channels.  We now construct an effective 1D model to derive its precise form


{\it The model} -- We consider a 1D Dirac channel along the $x$ direction $\hat{e}_x$ at $y=z=0$ described by the Hamiltonian
\be
\Hcal_0=-i\hbar v_F \partial_x \hat{\mv}\cdot\bsigma\,,
\ee
where $\hat{\mv}$ is a unit vector defining the spin-momentum locking direction (spin axis).  To generate a resistance we introduce a random impurity potential ${\mathcal V}(x)$
\be
{\mathcal V}(x) = V(x) - \frac{i\lambda^2}{2}\left\{ \nabla V \stackrel{\times}{,} \partial_x\hat{e}_x\right\}\cdot\bsigma\,,
\ee
where $\left\{ \,.\, , \,.\, \right\}$ denotes the anticommutator.  This includes a scalar potential $V(x) \equiv V(\br)|_{y=z=0}$ and a spin-orbit interaction term of the standard form.  The strength of the latter is controlled by the spin-orbit length $\lambda$.

\begin{figure}
\includegraphics[width=0.95\columnwidth]{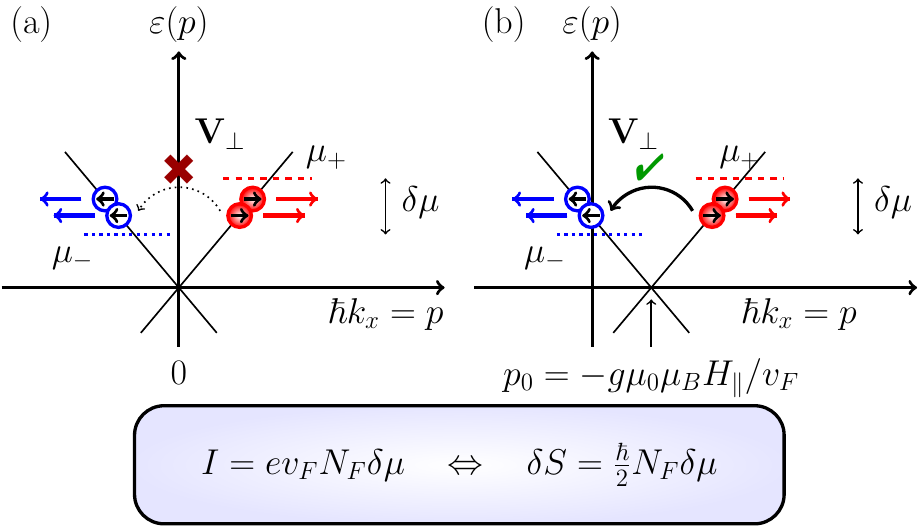}
\caption{(a) Energy dispersion for the model Hamiltonian $v_F p \sigma_x$, a special case of the spin-momentum locked Hamiltonian $v_F p \sigmav \cdot\hat\mv$.  The right movers with $p>0$ and positive spin along the $x$ axis have the same energy as their time-reversed left-moving partners with opposite momentum and spin. The short horizontal lines are the chemical potentials, which determine the occupations of right and left movers in the non-equilibrium state.  The elastic back scattering process shown by the arrow is forbidden by time-reversal symmetry.  (b): Energy dispersion for the model Hamiltonian $(v_F p+g\mu_B\mu_0 H) \sigma_x$.  Right and left moving states at the same energy do not have opposite momenta and spins, \ie they are not time-reversal partners. Elastic back scattering is allowed.  All electrons contributing to the linear-response current $I$ also contribute an $\hbar/2$ to the non-equilibrium spin polarization $\delta S$.}
\label{fig_Dirac_cones}
\end{figure}

It is convenient to separate  ${\mathcal V}(x)$ into components parallel $(V_\parallel)$ and orthogonal $(\bV_\perp)$ to the spin axis $\hat\mv$:
\be\label{ImpurityPotential}
{\mathcal V}(x)=  V(x) + V_\parallel\sigma_\parallel + \bV_\perp\cdot\bsigma_\perp.
\ee
Crucially, ${\mathcal V}(x)$ preserves time-reversal symmetry and therefore cannot cause elastic back-scattering in the absence of a magnetic field.  This is evident from the eigenstates of $\Hcal_0$: right $(+)$ and left $(-)$ movers $|\psi_p\pm\rangle$ are time-reversed partners, with $p=\hbar k_x$ the $x$-direction momentum, see Fig.\ref{fig_Dirac_cones}.  By Kramers' degeneracy, the matrix element $\langle\psi_{-p}\mp|{\mathcal V}(x)|\psi_p\pm\rangle$ vanishes.  Thus, applying a chemical potential difference $\delta \mu \equiv \mu_+ -\mu_-$ between reservoirs generates a charge current, as shown in Fig.~\ref{fig_Dirac_cones} for $\mu_+ > \mu_-$, but no backscattering within the channel, and only the contact resistance $R^{(0)}$ remains. 

 All this changes in presence of a magnetizing field $\Hv$, which modifies $\Hcal_0$ to 
\be
\Hcal_0 \to \left[v_F p + g\mu_0\mu_B \Hparallel\right]\sigma_\parallel+g\mu_0\mu_B\bH_\perp\cdot\bsigma_\perp,
\ee
with $g, \mu_0, \mu_B$ respectively the (effective) gyromagnetic ratio, vacuum permeability and Bohr magneton.  Regardless of its direction $\bH$ breaks time-reversal symmetry, but the parallel $(\Hparallel)$ and orthogonal $(\bH_\perp)$ components play qualitatively different roles. 

The field $\Hparallel$ merely shifts the dispersion in $k$-space, without creating any imbalance between right- and left-movers~\footnote{We assume that the system equilibrates to the state of minimum energy in the presence of the magnetizing field before any current is driven in the channel.}.  If the impurity scattering rate is assumed independent of $H$, such a shift can be dismissed as not contributing to the magneto-resistance~\cite{he2019}.  This is not the case here: shifted states at the same energy are no longer time-reversed partners, see Fig.~\ref{fig_Dirac_cones}, allowing back-scattering by ${\mathcal V}$.  Yet, in stark contrast to 2D scenarios, scalar ($V$) and parallel ($V_\parallel$) disorder components cannot scatter 1D Dirac states, even if shifted by $\Hparallel$, because counter-propagating states remain orthogonal in spin space.  Backscattering becomes possible only when transverse fields appear, $\bH_\perp\neq0, \bV_\perp\neq0$.  The resistance then increases according to the Landauer formula \cite{dattabook, mellobook}
\be
\label{eq_Landauer}
R= \frac{h}{e^2}\left[1+\frac{r}{t}\right]
\ee
where $r$ and $t$ are the reflection and transmission probabilities, and $t=1-r$.  Since $r$ must vanish for $\Hv=0$ and be even in $\Hv$, we have for small fields\footnote{To focus on the essentials without overburdening the notation we assume a diagonal $r_{ij}$.}
\be
\label{eq_ModelMR_1}
R= \frac{h}{e^2}\left[1+r_\perp H_\perp^2+r_\parallel H_\parallel^2\right]\,,\quad
r_\alpha=\frac{1}{2}\left.\frac{d^2r}{dH_\alpha^2}\right\vert_{H_\alpha=0}\,, 
\ee
with $\alpha=\perp, \parallel$.  We have so far ignored the exchange field arising from the current-induced spin polarization, an inevitable consequence of the spin-momentum-locked current.  Including this the Hamiltonian becomes
\ber
\label{eq_H_1}
\Hcal_0 &\to& \left[v_F p + g\mu_0\mu_B \Hparallel + J\delta \barS_\parallel\right]\sigma_\parallel+
\nn\\
&&\quad\quad +\left[g\mu_0\mu_B\bH_\perp+J\delta\barbS_\perp\right]\cdot\bsigma_\perp.
\eer
Here $\delta\barbS\equiv\delta\bS[I]/(\hbar/2)$ is the current-induced spin density in units of $\hbar/2$, linear in the applied bias, while $J$ is the exchange coupling constant, yielding the additional magnetizing field  $\bM[I]=J\delta\barbS[I]/g\mu_0\mu_B$.  Thus, the Hamiltonian \eqref{eq_H_1} contains  a mean-field non-equilibrium contribution from $\delta\bS$.  In the context of BMER the idea was proposed in Ref.~\cite{dyrdal2020}, though the physics of exchange coupling was neglected therein.  The exchange field generated by the current-induced spin polarization is however the dominant physical mechanism for the parallel Zeeman term in Eq.~\eqref{eq_H_1} -- the magnetostatic contribution is negligible \cite{SuppMat}.

Assuming the current-induced spin polarization to be entirely parallel to $\hat{\bf m}$ \footnote{Corrections in the form of orthogonal components may be generated by subleading mechanisms \cite{engel2007,gorini2017} which we neglect.}, the model simplifies. Only $H_\parallel$ contributes to BMER, and $H_\perp$ can be set to zero. For the same reason, neither the scalar impurity potential $V(x)$ nor the parallel component of the spin-orbit interaction $V_\parallel$ contribute to back-scattering under the stated assumptions.  The minimal model thus reads
\be
\label{eq_H_final}
{\mathcal H} = \left[v_F p + g\mu_0\mu_B \Hparallel + J\delta\barS_\parallel\right]\sigma_\parallel + \bV_\perp\cdot\bsigma_\perp\,,
\ee
where the spin density $\delta \barS_\parallel = \frac{I}{e v_F}$ is determined by spin-momentum locking, see Fig.~\ref{fig_Dirac_cones}, while the spin-orbit interaction with impurities reads 
\be
\bV_\perp\cdot\bsigma_\perp = \frac{\lambda^2}{2}\left[\left\{\partial_zV , -i\partial_x \right\}\sigma_y - \left\{\partial_yV , -i\partial_x \right\}\sigma_z\right]\,,
\ee
where we have assumed, without loss of generality, that $\hat\mv$ lies along $x$-axis.  The {\it transverse} random spin-orbit term is the sole ingredient capable of flipping the momentum-locked spin from right- to left movers.  This transverse random interaction is the central mechanism enabling BMER in 1D.  An established central feature of odd-in-$H$ resistance terms is indeed the necessity of some symmetry breaking in the direction transverse to transport \cite{sanchez2004, marlow2006}.  While this option is not available in 1D via orbital motion, it is achieved in spin space via $\bV_\perp$. The full BMER is obtained by substituting $\bH \to \bH + \bM$ in Eq.\eqref{eq_ModelMR_1}, giving 
\be
\label{eq_BMER_only1}
R_{\rm B} = \frac{h}{e^2} 2 r_\parallel M_\parallel H_\parallel\,,
\ee
where $M_\parallel=J\delta \bar S_\parallel/g\mu_0\mu_B$ is proportional to the current.
What is left is to compute the reflection coefficient $r_\parallel$. 


{\it The backscattering resistance} -- We sketch the calculation of $r_\parallel$, based on Fermi's Golden Rule. The perturbative approach is valid as long as backscattering is weak, $r_\parallel\ll1$, and the localization length of the 1D states exceeds the channel length $L$.  Details are given in Secs. 6-7 of the Supp. Mat. \cite{SuppMat}. 

To simplify the calculation, we apply the gauge transformation $U = e^{-i(\BB_\parallel x)/(\hbar v_F)}$  \cite{dyrdal2020}, where $\BB_\parallel\equiv g\mu_0\mu_B \Hparallel + J\delta \barS_\parallel$ is the effective parallel field.  This removes the Zeeman term from the Hamiltonian, leaving the transverse disorder ${\bV}_\perp\cdot\bsigma_\perp$ as the sole scattering potential.  Assuming standard $\delta$-correlated disorder, the impurity average yields $\overline{\partial_\alpha V(\br) \partial_\beta V(\br')}|_{\alpha=\beta=0} = n_i \varepsilon^2_0 \delta_{\alpha\beta}\delta(x - x')$, for $\alpha, \beta \in \left\{y, z\right\}$, where $n_i$ is the impurity concentration and $\varepsilon_0$ charaterizes the disorder strength.  The reflection probability $p\to-p$ in a channel of length $L$ is $r = (L/\hbar v_F)^2\overline{|\langle -p + |{\bf V}_\perp\cdot\bsigma_\perp|p + \rangle|^2}$.  Evaluating the matrix element one obtains
\be
R \approx \frac{e^2}{h}
  \left[
    1 + \left(\lambda k_F\right)^4 \left(\frac{\BB_\parallel}{\epsilon_F}\right)^2 N_i\left(\frac{\varepsilon_0}{\epsilon_F}\right)^2
  \right],
\ee
with $N_i = n_i L$ the total number of impurities.  In the $\BB_\parallel^2$ term we keep only the contribution linear in $H_\parallel$ and $M_\parallel$, yielding
\be
 \label{RBMER1}
 R_{\rm B} =
   \frac{h}{e^2}
   \left(\lambda k_F\right)^4 
   \frac{(g\mu_B\mu_0 H_\parallel)\,J\delta \bar{S}_\parallel[I]}{\epsilon_F^2}
   N_i\left(\frac{\varepsilon_0}{\epsilon_F}\right)^2.
\ee
This expression explicitly shows the dependence on the spin-orbit strength $(\lambda k_F)^4$, the disorder strength $(\varepsilon_0/\epsilon_F)^2$, and the product of the external field and current-induced polarization.

\begin{figure}
\includegraphics[width=0.95\columnwidth]{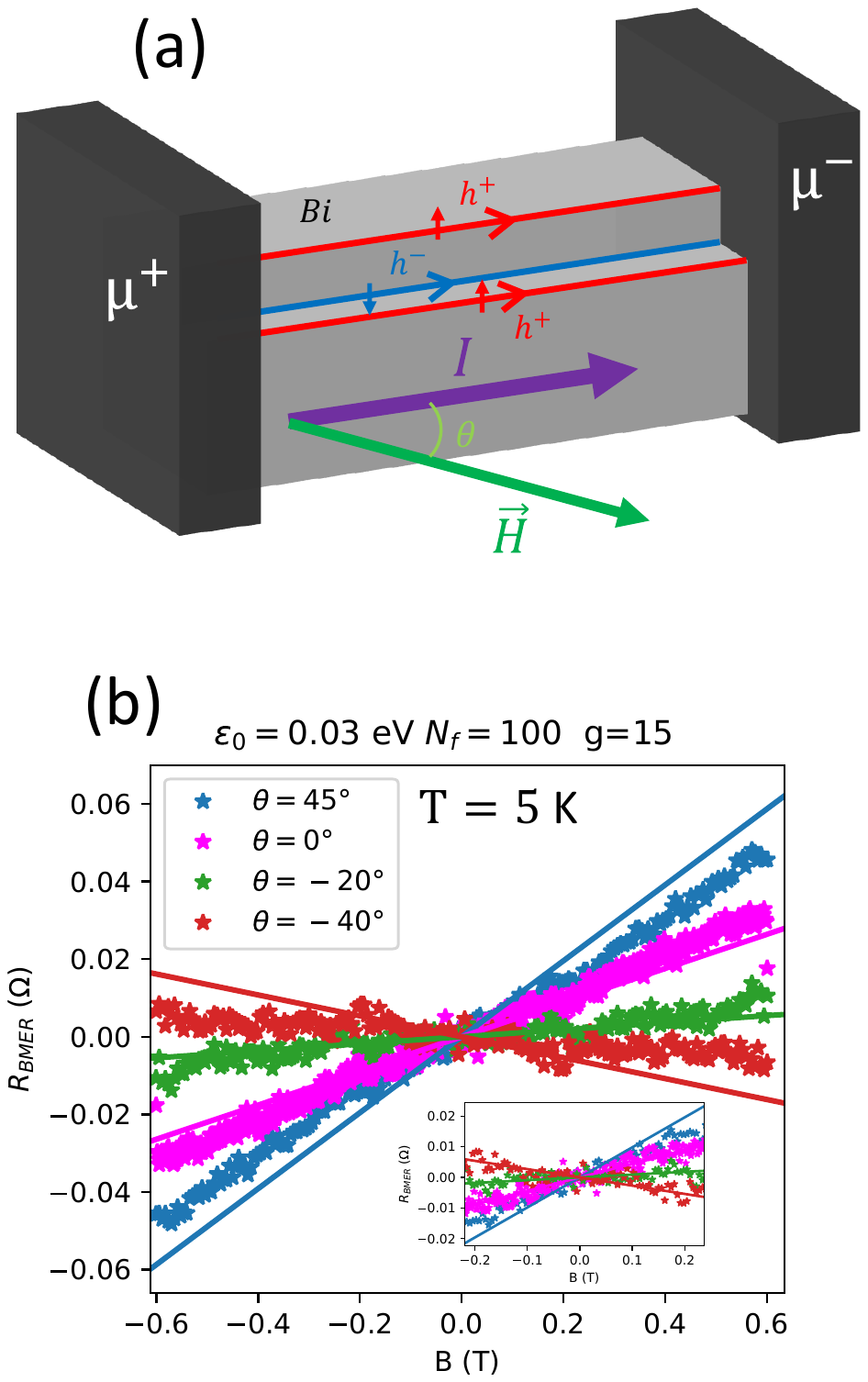}
\caption{{(a): Sketch of the experimental setup.  The shape of the Bi nanowire defines multiple hinges of either helicities, whose spin-momentum locking direction $\hat{\mv}$ forms an angle of $65^{\circ}$ with the nanowire axis. An external field is applied in the wire plane at an angle $\theta$.  The imposed total current is $I_{AC}=2.5 \mu{\rm A}$, through a wire of diameter $d \sim 100$nm.  See Ref.~\cite{bardpreprint} for further details. (b) BMER traces for varying $\theta$, corresponding to an angle between magnetic field and spin axis $\varphi = 65^{\circ}-\theta$.  The data were antisymmetrized to isolate the rectifying (2nd harmonic in current bias) odd-in-$B$ component, excluding $B$-even contributions.  Data points at small fields $B\lesssim 0.2$T, see inset, are perfectly fitted by the BMER theory, Eqs.~\eqref{RBMER1} - \eqref{eq_BMER_Bi_1}.
}}
\label{fig_exp}
\end{figure} 

{\it An explicit estimate} -- We use Eq.~(\ref{RBMER1})  to estimate BMER in Bi hinge states.  Fig.~\ref{fig_exp}(a) sketches the experimental setup: a Bi nanowire with $M>1$ hinges connecting two deposited metallic contacts, carrying a total DC current $I_{\rm tot}=2.5\mu$A. Support for the existence of such ballistic hinge channels comes from supercurrent measurements in similar Bi nanowires connected to superconducting electrodes. These experiments reveal a critical current modulated by a magnetic field with a periodicity corresponding to one flux quantum through the wire area, persisting up to several Teslas \cite{li2014, murani2017, bernard2023}. This behavior indicates that the supercurrent is confined to narrow, ballistic channels located on the hinges between crystal facets or along atomic step edges. Furthermore, the observation of a saw-tooth current-phase relation in these Josephson junctions confirms the ballistic nature of these 1D states.  Experimental data shows that the channels have the same spin axis $\hat{\bf m}$ \cite{bardpreprint}, but they are expected to have random helicities depending on which hinge they reside on -- compare red and blue states in Fig.~\ref{fig_exp}(a).  Copropagating channels with opposite helicity contribute to BMER with opposite signs. Let $M_+$ and $M_-$ be the numbers of copropagating positive- and negative-helicity channels, respectively.  In an ideal symmetric setup $M_+=M_-$, canceling the net current-induced spin polarization and thus the BMER.  In practice experimental imperfections (exact shape and position of the hinges, coupling to the electrodes) lead to a mismatch $\delta M = |M_+ - M_-| \neq 0$.  Simple calculations \cite{SuppMat} yield a {\it net} BMER given by
\be
\label{eq_BMER_Bi_1}
R^{\rm Bi}_{\rm B}  \approx  
\frac{\delta M}{M^2} R_{\rm B}\,.
\ee 
To estimate the exchange constant $J$, which controls the energy splitting between up and down spin states, we proceed by neglecting numerical factors of order 1.  On dimensional grounds, $J = \alpha \varepsilon/n_{1D}$, with $\alpha, \varepsilon, n_{1D}$ respectively an effective fine structure constant, a relevant energy scale and a 1D density.  The constant $\alpha$ is obtained by considering the typical Coulomb repulsion between Fermi surface electrons, $U_C$, and dividing it by the kinetic energy $\hbar k_F v_F$.  Since $U_C = e^2/(4\pi\epsilon_0\langle r\rangle) ~\approx e^2 k_F/(4\pi\epsilon_0)$, with $\epsilon_0$ the vacuum permittivity, one finds $\alpha \approx e^2/(4\pi\epsilon_0\hbar v_F) \approx \mathcal{O}(1)$.  The dependence on the Coulomb interaction is explicitly accounted for in $\alpha$, so the remaining $\varepsilon/n_{1D}$ must be a property of the non-interacting system.  The most plausible choice is $\varepsilon / n_{1D} = \epsilon_F / k_F = \hbar v_F$.  This leads to $J=\alpha \hbar v_F \simeq \hbar v_F$, and thus to 
\be
\label{eq_estimate}
J\delta \barS_\parallel \simeq \frac{h}{e}I \approx 2.581 \times 10^4 ({\rm eV/A})\, I.
\ee
This estimate makes topological fragility manifest: the {\it entire} current in a protected 1D state converts directly into a Zeeman splitting, which here does not open any gap.

Using parameters in the range relevant for Bi experiments \cite{bardpreprint}: External field $|B|$ in the range 0.1 - 1 T, with associated energy $|\mu_B B| \approx 5.788 \times 10^{-5}$ 0.1 - 1 eV; $M=100, \delta M = 10$, and a channel length $L=1\mu$m; Current per hinge $I = I_{\rm tot}/M = 0.025\mu$A, and thus a splitting $J\delta\bar{S}[I]\approx 0.65$meV; $\epsilon_F = 100$meV, $v_F = 0.5 \times 10^6$m/s, $g$ = 15; The critical quantity $\lambda$ is estimated by first recalling that the Elliott-Yafet spin flip rate $1/\tau_{EY}$ is proportional to the momentum scattering time $1/\tau\,$\cite{zutic2004}, and specifically $1/\tau_{EY}=(\lambda k_F/2)^4\,1/\tau\,$\cite{raimondi2010}, and then by noticing that measurements in Bi show $1/\tau_{EY} \approx 1/\tau\,$\cite{komnik2005}.  We thus have $(\lambda k_F/2)^4 \approx 1$.  The impurity number $N_i$ and the disorder strength $\varepsilon_0$ are the only free parameters in Eq.\eqref{RBMER1}, which reproduces the experimental BMER magnitude excellently for $N_i=100, \varepsilon_0/\epsilon_F \approx 0.3$.  These values are fully consistent with the known phenomenology of Bi, confirming that our minimal mechanism captures the essential physics.

{\it Outlook} -- We have demonstrated that non-reciprocal bilinear magnetoelectric resistance (BMER) in topologically protected 1D Dirac states is a robust, general phenomenon arising from ``topological fragility.'' The critical insight is that strong spin-orbit coupling plays a dual role: it is the very mechanism that stabilizes the topologically protected phase in the linear regime, yet its fluctuations due to ambient disorder simultaneously enable the Elliott-Yafet backscattering channel in the non-linear regime.

Unlike previous proposals invoking complex many-body effects or bulk coupling, our minimal model relies solely on the interplay of current-induced spin polarizations and ever-present disorder, and does not require any gap opening. This simple framework yields quantitative agreement with experimental BMER data from Bi hinge states, using parameters consistent with known Bi phenomenology.

Our work resolves the tension between observed non-reciprocal transport and the theoretical protection of 1D topological channels. Beyond explaining existing experiments, this mechanism predicts strong BMER in any 1D spin-momentum locked system with strong spin-orbit coupling, notably higher-order topological insulators. This opens new avenues for non-reciprocal spintronic devices and provides a unified understanding of transport in protected 1D channels, where the same interaction that guarantees protection ultimately enables its own violation.

{\it Acknowledgements} -- CG thanks the I-FIM, NUS for support and hospitality, and the STherQO members for useful discussions. GV acknowledges support from the Université Paris Saclay, LPS, and from the Ministry of Education, Singapore, under its Research Centre of Excellence award to the Institute for Functional Intelligent Materials (I-FIM, project No. EDUNC-33-18-279-V12). MB, SG and HB acknowledge discussions with Alexandre  Bernard,  Eli Gerber, Benjamin Wieder, Albert Fert, Manuel Bibes, Richard Deblock, Meydi Ferrier and Alexei Chepelianskii. The experimental part of this work was funded by the European Union through the BALLISTOP ERC 66566 advanced grant.  We also benefited from  technical help from S. Autier-Laurent and R. Weil at LPS.

\bibliographystyle{apsrev4-1}
\bibliography{BMER1D_biblio}


\section{End matter}

\appendix
\section{1. BMER à la Landauer-Büttiker}
\label{sec_Landauer-Buettiker}

Within the Landauer-Büttiker approach a systematic treatment of non-linear transport is possible \cite{buettiker1993,christen1996,sanchez2004}, but some generalizations are necessary to describe BMER in Dirac states.  For clarity's sake we present them within the context of a 1D two-terminal helical device, adopting the language of Refs.~\cite{buettiker1993,christen1996,gasparian1996}. The current as a function of the bias $V\equiv V_+ - V_-$ and magnetic field ${\bf B}$ reads
\be
I(V) = G_1({\bf B}) V + \frac{1}{2}G_2({\bf B}) V^2 + \dots.
\ee
The leading non-linearity $G_2$ is
\be
\label{eq_G2_basic}
G_2 = \frac{e^2}{h}\int\,{\rm d}\varepsilon (-f') 2\left[\partial_{V_+}-\partial_{V_-}\right] T(\varepsilon,V_+, V_-,{\bf B}),
\ee
with $f' = \partial_\varepsilon f$ the derivative of the Fermi function at $V=0$ and $T(\varepsilon,V_+, V_-,{\bf B})$ the screened transmission probability.  This expression must be generalized to
\be
\label{eq_G2_bmer}
\tilde{G}_2 = \frac{e^2}{h}\int\,{\rm d}\varepsilon (-f') 2\left[\partial_{V_+}-\partial_{V_-}\right] T(\varepsilon,V_+, V_-,\BB)
\ee
to describe BMER for Dirac electrons, the partial derivatives $\partial_{V_\pm}$ acting also on $\BB=\BB(V) = g\mu_B {\bf B} + J\delta\bar{\bS}(V)$.  

\subsection{The standard $G_2$ in 1D Dirac systems}
Each biased contact injects charges into the scattering region, locally adding $dn_i(x)/d\varepsilon$ charges per unit volume and energy, $i=\pm$.  Such ``injectivities'' add up to the full density of states $dn(x)/d\varepsilon = \sum_i dn_i(x)/d\varepsilon$ \cite{gasparian1996}.  Here the injectivities are local densities of right- ($i=+$) and left movers ($i=-$).  The potential in the scattering region is modified by such added charges, $U_0(x) \to U(x,V_+,V_-)$, and is computed solving Poisson equation with $dn_i(x)/d\varepsilon$ as sources.  Its symmetry thus follows from that of the injectivities, both without \cite{buettiker1993,christen1996,gasparian1996} and with magnetic fields \cite{sanchez2004}.  If one rewrites $G_2$ as
\be
\label{eq_G2_functional}
G_2 = \frac{e^2}{h} \int\,{\rm d}\br \frac{\delta T}{\delta U} 2\left[\partial_{V_+}-\partial_{V_-}\right]U,
\ee
clearly $G_2=0$ in a symmetric device [$U_0(x)=U_0(-x)$] with symmetric contacts [$dn_+(x)/d\varepsilon = dn_-(-x)/d\varepsilon$].  

\begin{figure}
\begin{center}
\includegraphics[width=\columnwidth]{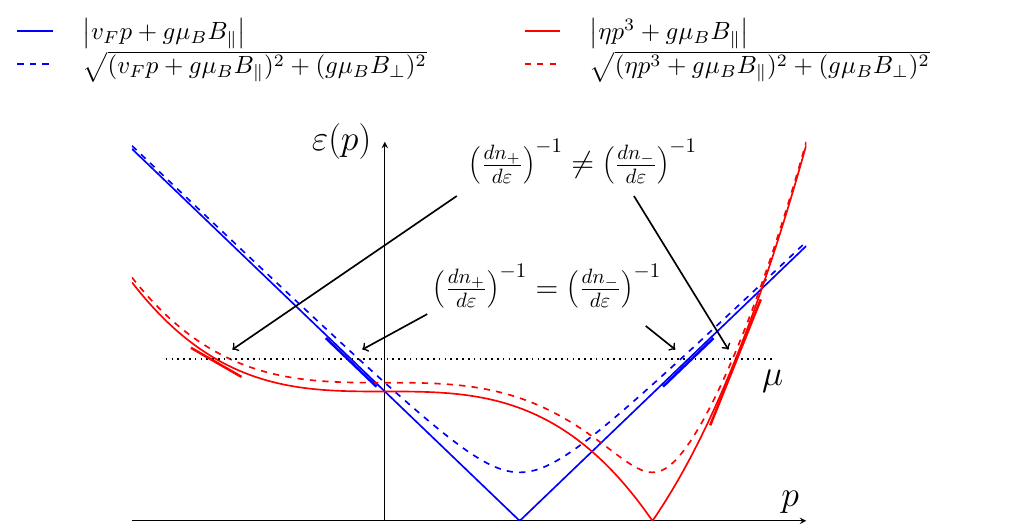}
\caption{Fermi level densities of injected right- (+) and left-movers (-) for linear (blue) and cubic (red) Dirac dispersions.}
\label{fig_end_matter}
\end{center}
\end{figure}

Consider the consequences for BMER in a symmetric device with a distorted Dirac Hamiltonian $\Hcal_0 = [{\bf b}({\bf p}) + g\mu_B{\bf B}]\cdot\bsigma$, where ${\bf b}({\bf p})=v_F{\bf p}+\eta{\bf g}({\bf p})$ and ${\bf g}({\bf p})$ is a nonlinear odd-in-${\bf p}$ function.  Its spectrum at ${\bf B}=0$ is such that right- and left-movers injectivities are identical, $dn_+(x,{\bf B}=0)/d\varepsilon = dn_-(-x,{\bf B}=0)/d\varepsilon$.  In presence of a magnetic field with finite parallel component, $B_\parallel\neq0$, this changes {\it if and only if} $\eta$ is non-vanishing, $dn_+(x,{\bf B})/d\varepsilon \neq dn_-(-x,{\bf B})/d\varepsilon \Leftrightarrow \eta \neq 0$, see Fig.~\ref{fig_end_matter}.  This proves that standard BMER, based on $G_2$, is absent in linearly-dispersing Dirac states, but may appear when cubic and higher distortions are present \cite{balram2019}.

\subsection{2. BMER from the generalized $\tilde{G}_2$}
Take the same symmetric Dirac device as in the previous Section.  The finite biases $V_+, V_-$ not only inject charges which modify the electrostatics, $U_0(x) \to U(x,V_+,V_-)$. They also induce a non-equilibrium spin polarization which modifies the magnetic field term in the Hamiltonian, $g\mu_B\bB\cdot\bsigma \to [g\mu_B\bB + J\delta\bar{\bS}(V)]\cdot\bsigma \equiv \BB(V_+, V_-)\cdot\bsigma$.  Thus, the transmission probability acquires a further bias dependence, $T(\varepsilon,V_+,V_-,\bB) \to T(\varepsilon,V_+,V_-,\BB(V_+,V_-))$.  From the simple chain rule, the leading non-linearity becomes
\ber
\tilde{G}_2 &=& \frac{e^2}{h} \int\,{\rm d}\br 
  \left\{
    \frac{\delta T}{\delta U} 2\left[\partial_{V_+}-\partial_{V_-}\right]U + \right.
    \nn\\
    && \hspace{2.5cm}
    \left. + \frac{\delta T}{\delta \BB} 2\left[\partial_{V_+}-\partial_{V_-}\right]\BB
  \right\}
  \nn
  \\
  &=&
  \frac{e^2}{h} 
  \left\{
      2\left[\partial_{V_+}-\partial_{V_-}\right]T|_{\epsilon_F, V=0} \,+ 
  \right.
      \nn
      \\
      &&
  \left.
        \hspace{.5cm} + 2 \partial_\BB T|_{\epsilon_F, V=0} \left[\partial_{V_+}-\partial_{V_-}\right]\BB|_{\epsilon_F, V=0}   
  \right\}.
\eer

We saw that without cubic corrections, \ie for $\eta=0$, the first term vanishes $\left[\partial_{V_+}-\partial_{V_-}\right]T|_{\epsilon_F, V=0}=0$.  The second term $\propto \partial_\BB T \partial_V \BB$ however does not, giving rise to BMER for linear 1D Dirac states as discussed in the main text.  By expanding the transmission probability at zero bias $T(\epsilon_F,0,0,\BB) \approx T(\epsilon_F,0,0,0) + \frac{1}{2}\frac{\partial^2T}{\partial\BB^2}  \, \BB^2$, and comparing with Eqs.~\eqref{eq_ModelMR_1} and \eqref{eq_BMER_only1} one finds
\be
-(g\mu_0\mu_B)^2 \frac{\partial^2 T}{\partial\BB^2} = r_\parallel.
\ee
The latter is the quantity that we explicitly computed to obtain the BMER expression~\eqref{RBMER1}.  Our general Landauer - Büttiker formulation allows to easily implement BMER calculations in standard numerical toolboxes for quantum transport, which typically compute the transmission function $T$.

\begin{widetext}
\section{Supplemental Material}
We provide derivations  of various results presented in the main paper.

\section{1. BMER in topological edge channels}

The edges of a non-magnetic topological insulator -- or specific hinges of higher-order topological insulators -- host two counter-propagating sets of states related to each other by a time reversal transformation.  In the absence of disorder, these are eigenstates of 1D momentum $k=\hbar/p$ with two-component spinor wave functions
\be
\psiv_{+k}(x,y,z)=e^{ikx}\phi(y,z) \wv(k)\,,~~~~\psiv_{-k}(x,y,z)=e^{-ikx}\phi(y,z) \tilde\wv(k)
\ee
Here  
\be
\tilde\wv(k) \equiv i\sigma_y  \wv^*(k)
\ee
 is the time-reversed partner of the two-component spinor $\wv(k)$, with $\sigma_y$ the 2nd Pauli matrix, and $\phi(y,z)$ is a real scalar function that ensures localization on the one-dimensional edge. 
The  two branches have identical energies 
\be
E_{+p}=E_{-p}\,,
\ee
not necessarily linear in $p$.  These energies are the positive eigenvalues of an effective Hamiltonian
\be
\Hcal(p)=\bb(p)\cdot\sigmav, \quad \sigmav = (\sigma_x, \sigma_y, \sigma_z),
\ee
where $\bb(p)$ is an odd vector-valued function of $p$ with $|\bb(p)|=E_{\pm p}$ and $\bb(p)=-\bb(-p)$ (for example $\bb(p)=v_Fp\hat\xv$ is the simplest choice). Notice that the vector of Pauli matrices $\hat\sigmav$ is dimensionless. 

It is easy to prove that the matrix element of any time-reversal-invariant impurity potential $V(x)$ between states at the same energy but propagating in opposite directions vanishes
\be
\langle \psiv_{-k}|V(x)|\psiv_{+k}\rangle=0.
\ee  
This strongly suggests that there is no back-scattering from  time-reversal-invariant (non-magnetic) impurities.  Indeed this is true at all orders in the impurity potential as the degeneracy of time-reversed partners is guaranteed by Kramers theorem.

If we now inject a charge current by creating an imbalance between the populations of right and left movers (\ie two different chemical potentials $\mu_+$ and $\mu_-$ for right and left movers respectively) we expect this current to be protected against back-scattering from non-magnetic impurities.  Since the back-scattering (= reflection) probability vanishes, the resistance of the channel is expected to be the pure contact resistance
\be
R^{(0)}=\frac{h}{e^2}.
\ee

When we apply a magnetizing field $\Hv = {\bf B}/\mu_0$ the effective hamiltonian changes to
\be
\hat \Hcal(p) = \left[\bb(p)+g\mu_0\mu_B\Hv\right]\cdot\hat\sigmav\,,
\ee
where $\mu_B$ is the Bohr magneton and $\mu_0$ the vacuum permeability.  Time reversal invariance is broken, therefore states at the same energy are no longer time-reversal partners of each other. The situation is illustrated in Fig. 1 for the simplest case $\bb(p)=\hbar v_Fp\hat\xv$, $\Hv=H\hat\xv$.  

\begin{figure}[ht]
\begin{center}
\includegraphics[width=12cm]{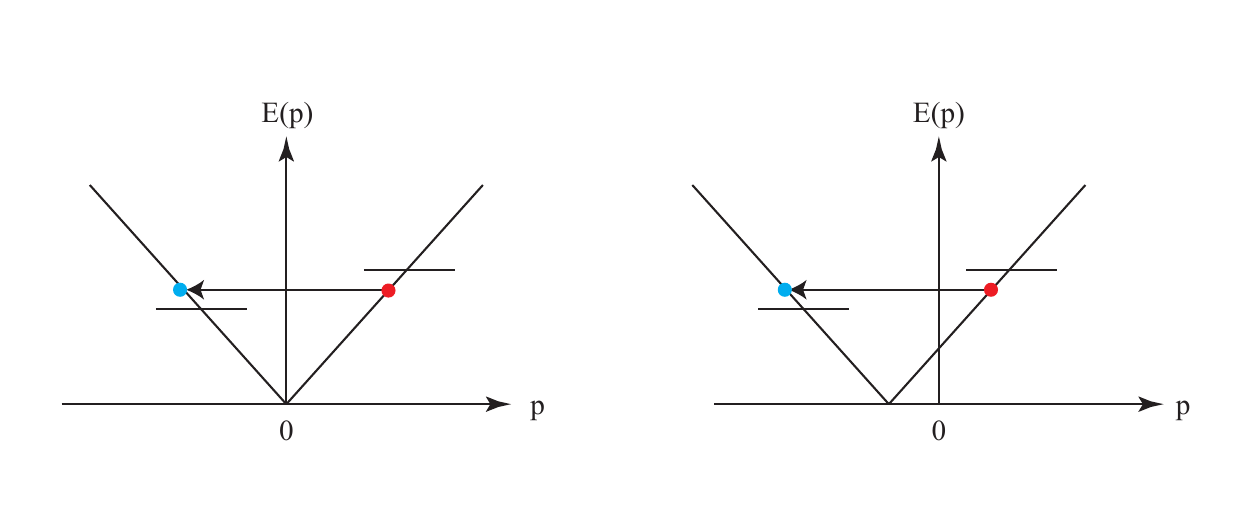}
\caption{Left panel: Energy dispersion for the model Hamiltonian $v_F p \sigma_x$.  The right movers with $p>0$ and positive spin along the $x$ axis have the same energy as their time-reversed left-moving partners with opposite momentum and spin.  The short horizontal lines are the chemical potentials, which determine the occupations of right and left movers in the non-equilibrium state.  The elastic back scattering process shown by the arrow is forbidden by time-reversal symmetry.  Right panel: Energy dispersion for the model Hamiltonian $(v_F p+g\mu_0\mu_B H) \sigma_x$.  Right- and left-moving states at the same energy do not have opposite momenta and spins, \ie they are not time-reversal partners. Elastic back scattering is allowed.}
\end{center}
\end{figure} 

This magnetizing field corresponds to $H_\parallel$ in the main text.  As discussed therein, $H_\parallel$ simply shifts the dispersion in momentum space.  If there was an imbalance between the occupations of right and left movers, this imbalance, as well as the current associated with it, is not changed by the shift.  The effects of the shift can be dismissed unless one recognizes that the scattering rate itself depends on $\Hv$: states at the same energy are not time-reversal partners of each other, therefore it becomes possible for a non-magnetic impurity to back-scatter an electron.  It is easy to see, as will be shown explicitly below, that the matrix element of the impurity potential between states at the same energy is non-zero if the non-magnetic impurity has some transverse component ${\bf V}_\perp$.  Another option is that the magnetizing field is such that $\Hv_\perp\neq0$.  Whatever the specific case, the channel resistance in general increases to
\be
\label{eq_ModelMR_11}
R = \frac{h}{e^2}\left[1+\frac{r}{t}\right] =
\frac{h}{e^2}\left[1+r_\perp H_\perp^2+r_\parallel H_\parallel^2\right]\,,\quad
r_\alpha=\frac{1}{2}\left.\frac{d^2r}{dH_\alpha^2}\right\vert_{H_\alpha=0}\,, 
\ee
with $\alpha=\perp, \parallel$, see arguments in the main text.

We have ignored so far the steady-state non-equilibrium magnetization of the channel ${\bf M}[I]$.  Its existence is however an inevitable consequence of spin-momentum locking and the current in the non-equilibrium state.  To be precise, the current-induced spin polarization $\delta {\bf S}[I]$ gives rise to a magneto-static correction to the Hamiltonian
\be
\delta {\cal H}_m = g\mu_B\mu_0{\bf H}_m\cdot\bsigma, 
\ee
where ${\bf H}_m$ is the field generated by ${\bf M}[I]$ within the nanowire, including demagnetization effects, see below.  In our simple model
\be
{\bf M} = g\mu_B\frac{\delta {\bf S}[I]}{\hbar}
\ee
with $\delta{\bf S} \parallel \hat{\bf m}$.  The overall field is thus the sum of the external and internal ones
\be
H_\parallel \to H_\parallel + H_{m\,\parallel},
\ee
yielding
\be
\label{eq_ModelMR_2}
R = \frac{h}{e^2}\left[1+r_\perp H_\perp^2+r_\parallel 
  \left(
    H_\parallel + H_{m\,\parallel}
  \right)^2
\right]
=
\frac{h}{e^2}\left[1 + r_\perp H_\perp^2 + r_\parallel(H^2 + 2 H_\parallel H_{m\,\parallel} + H_{m\,\parallel}^2)\right]\,.
\ee
The cross term $2 H_\parallel H_{m\,\parallel}$ is the origin of the magneto-static BMER.     

In the main text we developed an equivalent argument considering another option, namely that the non-equilibrium spin polarization is exchange-coupled to the electronic spins due to Coulomb interaction.  In general both mechanisms should be at work.  Below we estimate their relative strength.

\section{2. BMER: an explicit simple example}
\label{sec_simple_example}

As an example of the general BMER mechanism we consider the model introduced in Section 1.  A scalar impurity potential is unable to connect states at the same energy because they have opposite spin orientations.  However the spin can be flipped if we allow for  spin-orbit interaction between the impurity and the electron.  We place the charged impurity a distance $d$ below the $x$ axis -- at $x=0$ in the $(x,z)$ plane -- so the spin-orbit interaction is
\be
{\cal H}_{so} = \frac{\lambda^2 e}{\hbar} \left[p_x E_z(x)+E_z(x)p_x\right]\sigma_y\,,
\ee
where $\lambda$ is the effective Compton wavelength and
\be
E_z(x) = \frac{e}{4\pi \epsilon_0 d^2} \frac{1}{(1+x^2/d^2)^{3/2}}
\ee
is the $z$-component of the electric field created by the impurity on the edge channel. 
Notice that ${\cal H}_{so}$ is invariant under time reversal.

Now let us calculate the matrix element of ${\cal H}_{so}$ between states with wave vectors $k>0$ and $k'=-k-\mu_0H_\parallel/v_F$ (see Fig. 1).
The operator $\sigma_y$ takes care of flipping the spin from $+x$ to $-x$.  The remaining matrix element is
\ber
\langle k'|{\cal H}_{so}|k\rangle &=& \frac{\lambda^2}{L} e\int dx e^{-ik'x}[-i\partial_x E_z(x)+E_z(x)(-i\partial_x)]e^{ikx}\nn\\
&=&\frac{\lambda^2}{L} (k+k')\int dx e^{i(k-k')x}eE_z(x)\,,
\eer
where $L$ is the length of the 1D channel.
We notice first of all, that the matrix element vanishes when $k'=-k$ i.e. when the initial and final states are time-reversed partners.
Next, using the definitions of $k'$ and $E_z(x)$ and the integral
\be
\int dx e^{i(k-k')x}eE_z(x) = \frac{e^2}{2\pi \epsilon_0 d} (|k-k'|d)K_1(|k-k'|d)\,, 
\ee
where $K_1(x)$ is the Bessel function BesselK, we obtain
\be
\langle k'|{\cal H}_{so}|k\rangle = -\frac{\lambda^2}{L} \frac{\mu_0 \mu_B H_\parallel}{\hbar v_F} \frac{e^2}{2\pi \epsilon_0 d} (|k-k'|d)  K_1(|k-k'|d)\,.
\ee

To leading order in $H$ we can set $k'=-k$ so we finally have
\be
\langle -k|{\cal H}_{so}|k\rangle \simeq -\frac{\lambda^2}{L}\frac{\mu_0 \mu_B H_\parallel}{\hbar v_F}  \frac{e^2}{\pi \epsilon_0 d}(kd)  K_1(2kd)
\ee

The reflection coefficient is given by
\be
r=|N(0)\langle -k|{\cal H}_{so}|k\rangle|^2
\ee
where 
\be
N(0) = \frac{L k_F}{\epsilon_F}
\ee
is the density of states  at the Fermi level. $k_F$ is the Fermi wave vector.
Thus we have
\be
r = \left\vert  (k_F \lambda)^2\frac{\mu_0 \mu_B H_\parallel}{\epsilon_F}  \frac{e^2}{\pi \epsilon_0 d \epsilon_F}(k_Fd)  K_1(2k_Fd)\right\vert^2 \,,
\ee
which is dimensionless as it should.  The coefficient $r_\parallel$ appearing in Eqs.~\eqref{eq_ModelMR_1} and \eqref{eq_ModelMR_2} follows immediately and determines the BMER.

\subsection{3. Edge channel magneto-static field from spin-momentum locking}

We model the edge channel as a cylinder of radius $a$ infinitely extended along the $z$ axis.  The magnetization density is uniform within the cylinder and zero outside:
\be\label{MagnetizationDensity2}
\Mv(\rho)=\frac{g\mu_B \delta\bar{S}}{\pi a^2}\Theta(a-\rho)\hat \mv
\ee
where $\delta \bar{S}$ is the uniform linear spin density along the channel in units of $\hbar/2$, see main text, $\rho$ the distance from the axis of the cylinder, $\Theta$ the step function, $\hat \mv$ the (fixed) unit vector in the direction of the spin axis.

From Maxwell's equations
\be
\nabla \times {\bf B} = \mu_0 {\bf j} = \mu_0 \left(\nabla \times {\bf M}\right) 
\ee
one has
\be
{\bf B} = \mu_0 \left({\bf M} - \nabla u\right) = \mu_0 ({\bf M} + \tilde{\bf H}).
\ee
Here $u$ is the magnetic scalar potential and $\tilde{\bf H}$ the demagnetizing field.  Such a field opposes ${\bf M}$ and is due to boundary effects.  

We have an effectively infinite wire along $z$, so that our problem becomes independent of $z$, $u \to u(\rho,\varphi)$.  We further decompose
\be
{\bf M} = {\bf M}_\perp + M_z \hat{\bf e}_z.
\ee
By symmetry, only ${\bf M}_\perp$ will generate a demagnetising field.  In magnitude $M_\perp = M\sin\vartheta$, with $\vartheta$ the angle between $\mv$ and the $z$-axis.

One has
\be
\nabla\cdot{\bf B} = 0 \Rightarrow \nabla\cdot{\bf M} = u'',
\ee
with $u'' = \partial^2_\rho u$.  In the outer region $\rho > a$ there is no magnetisation, while in the inner region $\rho < a$ the magnetisation is homogeneous, ergo in both cases the equation for $u$ is homogeneous
\be
u'' = 0.
\ee 
We require $u$ to vanish at both infinity and zero, therefore
\ber
\label{eq_gin1}
u_< &=& f_0 (\varphi) + f_1(\varphi)\rho = f_1(\varphi)\rho
\\
\label{eq_gout1}
u_> &=& h_0 (\varphi) + h_1(\varphi)\frac{1}{\rho} = h_1(\varphi)\frac{1}{\rho},
\eer
with $u_< (u_>)$ the potential in the inner (outer) region.  The $\varphi$-dependent coefficients can be expanded in harmonics
\be
f_1(\varphi) = \sum_n c_n \cos(n\varphi)
,\quad
h_1(\varphi) = \sum_n d_n \cos(n\varphi). 
\ee
To proceed, one considers the surface discontinuity (effective magnetic surface charge density) 
\be
\label{eq_jump1}
\sigma_m = {\bf M}\cdot\hat{e}_\rho = M_\perp \cos\phi = u'|_{a_-} - u'|_{a_+}, 
\ee
where $a_\pm = a \pm \delta$, with $\delta \to 0$.  Clearly, only the $n=1$ harmonic contributes
\be
f_1(\varphi) = c_1 \cos\varphi
,\quad
h_1(\varphi) = d_1 \cos\varphi,
\ee
so that
\ber
\label{eq_gin2}
u_< &=& c_1 \rho \cos\varphi
\\
\label{eq_gout2}
u_> &=& h_1 \cos\varphi \frac{1}{\rho}.
\eer
Imposing continuity $u_<(a) = u_>(a)$ one obtains
\ber
\label{eq_gin3}
u_< &=& c \rho \cos\varphi
\\
\label{eq_gout3}
u_> &=& c \cos\varphi \frac{a^2}{\rho}.
\eer
The constant $c$ is fixed by the jump condition Eq.~\eqref{eq_jump1}
\be
M_\perp\cos\varphi = c \left[1 + \frac{a^2}{r^2}|_a \right] \cos\varphi = 2 c \cos\varphi,
\ee
yielding
\be
\tilde{H}_< = -u'_< = -\frac{M_\perp}{2}\cos\varphi.
\ee
We conclude
\be
{\bf B}_< = \mu_0 \frac{{\bf M}_\perp}{2} + \mu_0 M_z \hat{\bf e}_z.
\ee

\subsection{4. Exchange versus magnetostatics}
From the discussion above and in the main text there are two non-equilibrium corrections to the edge/hinge channel Hamiltonian, one from exchange, $\delta{\cal H}_x$, and one from magneto-statics, $\delta{\cal H}_m$:
\be
\delta{\cal H}_x + \delta{\cal H}_m = J\delta\bar{\bf S}\cdot\bsigma + g\mu_B{\bf B}_<\cdot\bsigma.
\ee
The corresponding energies are
\be
E_x=J\delta\bar S\,,~~~~~~J \simeq\hbar v_F
\ee
and
\be
E_m=\mu_0 \frac{(g\mu_B)^2 \delta \bar S}{\pi a^2}\,.
\ee
Their ratio is
\be
\epsilon\equiv\frac{E_m}{E_x} =g^2\frac{\mu_0}{\pi} \frac{\mu_B^2}{\hbar v_F a^2}\,.
\ee
With $\frac{\mu_0}{\pi}=4 \times 10^{-7} {\rm N}/{\rm A}^2$, $\mu_B= 9.27 \times 10^{-24} {\rm Am}^2$, $\hbar = 6,63 \times 10^{-34}$Js, $v_F = 10^{-6}$m/s,$a=\alpha 10^{-9}$m, one has
\be
\epsilon \simeq 51 \frac{10^{-7} \, 10^{-48}}{10^6 \, 10^{-34} \, 10^{-18}} \frac{g^2}{\alpha^2}
 \simeq 5 \, \frac{g^2}{\alpha^2} 10^{-8}
\ee 
Typically $1 \lesssim \alpha \lesssim 10$, $1 \lesssim g \lesssim 20$, so that exchange strongly dominates the magneto-static contribution.  This would hold even in very narrow edge channels deep in the bulk gap, $\alpha \approx 1$, with large effective $g \approx 10$.  The two contributions could be comparable only if Coulomb interaction is heavily screened, possibly in presence of nearby gates.

\subsection{5. Remarks on the ``non-equilibrium spin-orbit field''}

For completeness, let us mention a subtle issue related to the definition of a current-induced non-equilibrium correction to the Hamiltonian.

In the literature \cite{dyrdal2020,vaz2020} it was proposed to consider a correction due to the spin-orbit field averaged over the non-equilibrium state.  Namely, from the standard spin-orbit Hamiltonian
\be
\Hcal_{so, 0} = {\bf b}(\bp)\cdot\bsigma,\quad \langle {\bf b}(\bp) \rangle_0 = 0,
\ee
it was proposed to obtain the BMER by taking
\be
\Hcal_{so, 0}+ H_{neq} = {\bf b}(\bp)\cdot\sigma + \langle {\bf b} \rangle_{neq}\cdot\bsigma.
\ee 
Here $\langle {\bf b} \rangle_{neq}$ is the spin-orbit field averaged over the current-carrying state.  The suggestive rewriting in terms of an ``exchange constant'' ${\mathcal J}$ was also put forward
\be
\langle {\bf b} \rangle_{neq} = {\mathcal J} {\bf S}\cdot\bsigma.
\ee
Yet, the constant ${\mathcal J}$ appearing here has nothing to do with Coulomb-driven exchange physics.  Interestingly, this was recognized already in Ref.~\cite{vaz2020}, where a warning was issued in a footnote.

\section{6. Fermi Golden Rule in a minimal model}
\label{sec_Fermi_min}

Take the model Hamiltonian
\be
\label{Hamilton_1}
\Hcal = \hbar v_Fp_x\sigma_\parallel+\BB\cdot\bsigma + 
\underbrace{V_0 + V_\parallel\sigma_\parallel + {\bf V}_\perp\cdot\bsigma_\perp}_{\delta\Hcal},
\ee
with the established notation where ``$\parallel$'' is defined by the spin-momentum locking texture, whatever this is for the 1D topological channel, while the directions orthogonal to it are denoted by ``$\perp$''.   

The field $\BB$ has the dimensions of an energy, and contains also the non-equilibrium steady-state component $\bDelta[I]$ from current-induced spin polarisation
\be
\label{eff_field}
\BB = g\mu_B\mu_0{\bf H} + \bDelta = g\mu_B{\bf B} + \bDelta,
\ee
whether from exchange or magneto-statics.

The part $\delta H$ describes disorder:
\begin{itemize}  
\item The term $V_0$ comes from standard impurities.  This can cause backscattering only if time-reversal is broken by someone else.
\item The term with spin structure $V_\parallel$ is not critical for BMER: being ``$\parallel$'' it does not cause any additional backscattering, as it cannot mix the eigenstates.
\item The term(s) $V_\perp$ are crucial: without them BMER as discussed here is not possible.  Indeed, such terms are very general and come from random spin-orbit coupling, as explicitly introduced in Sec.~\ref{sec_simple_example} for a single impurity close to the edge/hinge. 
\footnote{Note that the ``random Dirac cone'' model from Ref.~\cite{dyrdal2020} is fundamentally different, though still related to some randomness in spin-orbit coupling.  Indeed, such model does not work for 1D edge states, yielding only a $V_\parallel$ component, {\it ergo} no BMER.}
\end{itemize}
All potential terms are multiplied by an $e$ to have energy dimensions.

\subsection{7. BMER from random spin-orbit interaction}
\label{subsec_explicit}

As a specific form of \eqref{Hamilton_1} we take ($p_x \to -i\hbar\nabla_x$)
\be
\label{Hamilton_2}
\Hcal = \underbrace{-i \hbar v_F \nabla_x \sigma_x + \BB_x \sigma_x}_{\parallel} + \BB_\perp\cdot\bsigma_\perp + \underbrace{
\frac{\lambda^2}{2}\left[\left\{\partial_zV , -i\nabla_x \right\}\sigma_y + \left\{\partial_yV , -i\nabla_x\right\}\sigma_z\right]
}_{\delta V_\perp}.
\ee
The random part $\delta V_\perp\equiv {\bf V}_\perp \cdot \bsigma_\perp$ comes from SOC with the disorder potential $V(\br)$
\be
\frac{\lambda^2}{2} \left\{\nabla V \stackrel{\times}{,} -i\nabla_x\bhx \right\}\cdot \bsigma,
\ee
with $\lambda$ a material-dependent effective Compton wavelength.

With the gauge transformation
\be
U = e^{-i\frac{\BB_\parallel x}{\hbar v_F}}
\ee
the Hamiltonian \eqref{Hamilton_2} becomes 
\be
H' = \underbrace{-i\hbar v_F \nabla_x \sigma_x + \BB_\perp\cdot\bsigma_\perp}_{H_0'} +
\frac{\lambda^2}{2}\left[\left\{\partial_zV , -i\nabla_x - \frac{\BB_\parallel}{\hbar v_F}\right\}\sigma_y + \left\{\partial_yV , -i\nabla_x - \frac{\BB_\parallel}{\hbar v_F}\right\}\sigma_z\right].
\ee
The part $H_0'$ can be diagonalised
\be
\epsilon_\pm = \pm \sqrt{(v_Fp)^2 + \BB_\perp^2} \,; 
\;
\langle x|k\pm\rangle = \frac{1}{\sqrt{2L}}\frac{e^{ikx}}{\sqrt{1+|c_k^\pm|^2}}
\left(
 \begin{array}{c}
  1 \\
  c_k^\pm
 \end{array}
\right);\;
c_k^\pm = \frac{\epsilon_\pm - \BB_z}{\hbar v_F k -i \BB_y},
\ee
where $\nu\equiv\pm$ denotes the upper/lower 1D Dirac cone.  The random part is instead averaged assuming the potential to be isotropic and short range
\be
\overline{\partial_zV(\br) \partial_zV(\br')}|_{y=z=0} = \overline{\partial_yV(\br) \partial_yV(\br')}|_{y=z=0} = n_i \varepsilon^2_0 \delta(x - x'),
\ee
where the fluctuations $\varepsilon^2_0$ have the dimension of an energy squared [(eV)$^2$], while $n_i\equiv N_i/L$ is the impurity density along the 1D channel of length $L$. A more detailed disorder model can be considered, but it is irrelevant now.

In momentum space one has
\ber
 \langle k'\nu' |\delta V_\perp|k\nu\rangle &=&
   \frac{\lambda^2}{2L}\int\,{\rm d}x\,e^{-ik'x} 
   \left[
   \left\{
    \partial_zV,-i\partial_x - \frac{\BB_\parallel}{\hbar v_F}
   \right\}\langle\sigma_y\rangle +
   \left\{
    \partial_yV,-i\partial_x - \frac{\BB_\parallel}{\hbar v_F}
   \right\}\langle\sigma_z\rangle
   \right] e^{ikx}
   \nn\\
   &=&
   \frac{\lambda^2}{2L}\int\,{\rm d}x\, 
   \left[
    \left(k'-\frac{\BB_\parallel}{\hbar v_F}\right)\partial_zV + i\partial_x\partial_zV - i\partial_x\partial_zV + \left(k-\frac{\BB_\parallel}{\hbar v_F}\right)\partial_zV
   \right]
   e^{-i(k'-k)x}\langle\sigma_y\rangle +
   \nn\\
   && \quad\quad\quad
   + \left[
    \left(k'-\frac{\BB_\parallel}{\hbar v_F}\right)\partial_yV + i\partial_x\partial_yV - i\partial_x\partial_yV + \left(k-\frac{\BB_\parallel}{\hbar v_F}\right)\partial_yV
   \right] e^{-i(k'-k)x}\langle\sigma_z\rangle
   \nn\\
   &=&
   \frac{\lambda^2}{2L}\int\,{\rm d}x\, 
    \left(k'+k - \frac{2\BB_\parallel}{\hbar v_F}\right)
    \left(\partial_zV\langle\sigma_y\rangle + \partial_yV\langle\sigma_z\rangle\right)e^{-i(k'-k)x},
\eer
having performed an integration by part in the first passage, and with $\langle\sigma_i\rangle, i=y, z$ shorthand for $\langle\nu'|\sigma_i|\nu\rangle$.  There follows
\ber
\overline{|\langle k'\nu' |\delta V_\perp|k \nu \rangle|^2} &=&
 \left[\frac{\lambda^2}{2L}\left(k'+k-\frac{2\BB_\parallel}{\hbar v_F}\right)\right]^2
  \iint\,{\rm d}x {\rm d}x'\,
   \left(\overline{\partial_zV(x)\partial_zV(x')}|\langle\sigma_y\rangle|^2 + \dots \right)e^{-i(k'-k)(x-x')}
\nn\\
 &=&
 \left[\frac{\lambda^2}{2L}\left(k'+k-\frac{2\BB_\parallel}{\hbar v_F}\right)\right]^2
 \iint\,{\rm d}x {\rm d}x'\, n_i \varepsilon^2_0\delta(x-x')\left(|\langle\sigma_y\rangle + \langle\sigma_z\rangle|^2\right)e^{-i(k'-k)(x-x')}
\nn\\
\label{eq_deltaV2_avg}
 &=&
  \left[\frac{\lambda^2}{2L}\left(k'+k-\frac{2\BB_\parallel}{\hbar v_F}\right)\right]^2 (n_i L)\,\varepsilon^2_0 \left(|\langle\sigma_y\rangle + \langle\sigma_z\rangle|^2\right).
\eer

The Fermi Golden Rule backscattering rate $k\to k'=-k$ for a single hinge, and for electrons with a Fermi energy in the upper band, $\nu=+$, is
\be
\Gamma_k = \frac{L}{\hbar^2} N_{-k} \overline{|\langle -k + |\delta V_\perp|k + \rangle|^2},\quad N_{-k}=1/v_F
\ee
with corresponding reflection probability
\be
\label{eq_backscattering_r}
r = \left(\frac{L}{\hbar v_F}\right)^2\overline{|\langle -k + |\delta V_\perp|k + \rangle|^2}.
\ee
All quantities are evaluated at the Fermi energy $\epsilon_F$.  Inserting \eqref{eq_deltaV2_avg} into \eqref{eq_backscattering_r} one gets
\be
r = \left(\lambda k_F\right)^4 \left(\frac{\BB_\parallel} C {\epsilon_F}\right)^2 N_i\left(\frac{\varepsilon_0}{\epsilon_F}\right)^2 \approx \left(\lambda k_F\right)^4 \left(\frac{\BB_\parallel}{\epsilon_F}\right)^2 N_i\left(\frac{\varepsilon_0}{\epsilon_F}\right)^2,
\ee
where the factor $|\langle\sigma_y\rangle + \langle\sigma_z\rangle|^2 \equiv C \sim \mathcal{O}(1)$ was set to 1 in the last passage, and with  
\ber
\BB_\parallel^2 &=& \left(g\mu_B B_x\right)^2 + g\mu_B B_x \Delta_x + \Delta_x^2
\nn\\
 &=& \left(g\mu_B B_x\right)^2 + (g\mu_B B_x) J\delta\bar{S}[I] + \mathcal{O}(E^2),
\eer
considering the exchange contribution $\Delta_x = J\delta \bar{S}[I]$.  The BMER $\delta R$ comes from the linear-in-$I$ (or linear-in-$E$) term, that is
\be
 \label{eq_BMER_single}
 \delta R =
   \frac{h}{e^2}
   \left(\lambda k_F\right)^4 \left(\frac{(g\mu_B B_x)\,J\delta \bar{S}[I]}{\epsilon_F^2}\right) N_i\left(\frac{\varepsilon_0}{\epsilon_F}\right)^2.
\ee
This is the BMER resistance for a single hinge.  Below we consider the more realistic situation in which the current $I$ flows through a series of hinges.

\section{8. Combining different helicities}
In a real sample there will be $M>1$ physical edges transporting in parallel, in principle each with slightly different properties such as Fermi energy and velocity.  Neglecting such variations we assume all edges to be identical in this respect.  One has 
\be
\frac{1}{{\mathcal R}} = \frac{M}{R},
\ee
with ${\mathcal R}$ the full resistance, and that the full current $I_{\rm tot}$ equally redistributes through the $M$ channels
\be
I = \frac{I_{\rm tot}}{M}.
\ee
Experimental data shows that current-carrying states have identical spin quantization axis.  They may however come in two flavours
\be
M = M_+ + M_-
\ee
with $M_\pm$ the positive- and negative-helicity channels, respectively.  Helicity determines the sign of the current-induced spin polarization $\delta S$ carried by each channel, {\it ergo} the sign of $R_{\rm B}$.  In a quantum spin Hall sample each positive-helicity state residing on a given edge has a negative-helicity partner on the opposite edge.  Here however the helicity depends on how the crystal terminates at the specific hinge.  Furthermore not all hinges are equally coupled to the metallic electrodes.  Thus, in general one expects $M_+ \neq M_-$.  If helicity is random, this should however average to $M_+=M_-$ in the large $M$ limit.  

Focusing on the essentials, we assume that, for whatever sample-dependent reasons, we can include all randomness in the single parameter $\delta M \equiv M_+ - M_- \neq 0$.  Then, denoting by $R_+$ the combined resistance of the positive chirality channels and by $R_-$ the combined resistance of the negative chirality channels, one has for the total resistance
\be\label{TotalR}
\cR=\frac{R_+ R_-}{R_++R_-}\,.
\ee
Now let us write
\be
R_+ =\frac{1}{M_+}(R + \delta R)
\ee
and
\be
R_-=\frac{1}{M_-}(R - \delta R)
\ee
where $\delta R$ is the BMER correction which has opposite signs for opposite helicities.
Putting  $R_+$ and $R_-$ in Eq.~(\ref{TotalR}) we get
\ber
\cR &=& \frac{R^2-(\delta R)^2}{M_-(R+\delta R)+ M_+(R-\delta R)}\nn\\
&\simeq& \frac{R}{M} \frac{1}{1-\frac{M_+-M_-}{M} \frac{\delta R}{R}}\nn\\
&\simeq&\frac{R}{M} +\frac{M_+-M_-}{M^2}\delta R
\nn\\
&=&\frac{R}{M} +\frac{\delta M}{M^2}\delta R.
\eer
From this we see that the  total BMER of the system is given by
\be
\label{eq_BMER_M0}
\cR_{\rm B}= \frac{\delta M}{M} \frac{\delta R}{M}\,.
\ee
Notice that $\frac{\delta R}{M}$ is the BMER that one would obtain considering $M$ identical channels of a given (say positive) helicity. We see that the correct result for a mixture of positive and negative helicity channels is obtained by multiplying such result by the factor
\be
\frac{\delta M}{M}\,.
\ee
This factor can be as small as $1/100$ if $M \sim 100$ and $M_+-M_-$ is of order $1$, corresponding to full white-noise randomness of the variable $\delta M$.

Inserting Eq.~\eqref{eq_BMER_single} into Eq.~\eqref{eq_BMER_M0} one obtains
\be
\label{eq_BMER_M1}
\cR_{\rm B} = \frac{\delta M}{M^2}\frac{h}{e^2}
   \left(\lambda k_F\right)^4 \left(\frac{(g\mu_B B_x)\,J\delta \bar{S}[I]}{\epsilon_F}\right)^2 N_i\left(\frac{\varepsilon_0}{\epsilon_F}\right)^2,
\ee
with $I=I_{\rm tot}/M$.  The BMER estimates in the main text use this general formula.

\commentout{
\subsection{9. On randomness and symmetry-breaking}

From a macroscopic perspective, in a perfectly inversion-symmetric system the overall current-induced spin polarization must vanish.  In practice symmetry is broken by various mechanisms, \eg disorder, gates, sample shape, contacts.

For random symmetry-breaking, \eg from disorder, in large samples (self-averaging) one expects zero net current-induced spin polarization.  This is not the case in smaller ones, as considered above. 

However it is also possible to have systems that {\it on average} break inversion symmetry along a certain axis.  Simple mechanisms of this kind likely relevant for setups of the Bi-type are:
\begin{enumerate}
    \item The layered sample structure, composed of a series of plateaus and cliffs where hinge states reside.  When metallic contacts are deposited on such a symmetry-broken structure, hinges at the top of cliffs may have better contact with the electrodes than those at the bottom;
    \item Back- or top-gating;
    \item Doping and/or strain from the substrate.
\end{enumerate}
}

\end{widetext}

\end{document}